\documentclass[sigconf]{acmart}
\usepackage[T1]{fontenc}
\usepackage{algorithm}  
\usepackage{algpseudocode}  
\usepackage{amsmath} 
\usepackage{booktabs}
\AtBeginDocument{%
	\providecommand\BibTeX{{%
			\normalfont B\kern-0.5em{\scshape i\kern-0.25em b}\kern-0.8em\TeX}}}

\copyrightyear{2021}
\acmYear{2021}
\setcopyright{acmlicensed}\acmConference[BIOKDD '21]{20th International Workshop on Data Mining in Bioinformatics
}{August 14-18, 2021}{Virtual Conference}
\acmBooktitle{Proceedings 20th International Workshop on Data Mining in Bioinformatics (BIOKDD '21), October August 14-18, 2021, Virtual Conference}
\acmPrice{15.00}
\acmDOI{}
\acmISBN{}

\begin{document}
	\fancyhead{}
	\title{Improving Ultrasound Tongue Image Reconstruction from Lip Images Using Self-supervised Learning and Attention Mechanism}
	
	\author{Haiyang Liu}
\affiliation{%
	\institution{Graduate School of Information Science and Technology
	The University of Tokyo}
	\city{Tokyo}
	\country{Japan}}
\email{liuhaiyang@kmj.iis.u-tokyo.ac.jp}

\author{Jihan Zhang*}

\affiliation{%
	\institution{School of Mechanical Engineering\\
	Southeast University}
	\city{Nanjing}
	\country{China}}
\email{florentino_zhang@fuji.waseda.jp}
\thanks{* Haiyang Liu and Jihan Zhang contribute equally to this paper}	
	
	\begin{abstract}
	Speech production is a dynamic procedure, which involved multi human organs including the tongue, jaw and lips. Modeling the dynamics of the vocal tract deformation is a fundamental problem to understand the speech, which is the most common way for human daily communication. Researchers employ several sensory streams to describe the process simultaneously, which are incontrovertibly statistically related to other streams. In this paper, we address the following question: given an observable image sequences of lips, can we picture the corresponding tongue motion.
	We formulated this problem as the self-supervised learning problem, and employ the two stream convolutional network and long-short memory network for the learning task, with the attention mechanism.
	We evaluate the performance of the proposed method by leveraging the unlabeled lip videos to predict an upcoming ultrasound tongue image sequence. The results show that our model is able to generate images that close to the real ultrasound tongue images, and results in the matching between two imaging modalities.
	\end{abstract}

	
	
	\maketitle
	
	\section{Introduction}
	
	In the natural speech production process, researchers have employed a number of sensory streams to describe the highly variable movements of articulators simultaneously \cite{xu2020predicting}. Accurately modeling the deformation of vocal tract not only can be helpful to the theoretical quest in speech-related researches but also can be helpful for lots of seen applications, such as pronunciation training \cite{ding2019golden}, speech therapy and biosignal-based spoken communication \cite{schultz2017biosignal} (including: Silent Speech Interfaces (SSI) \cite{denby2010silent}).
	
	Since last several decades, different imaging techniques have been employed to analyze the speech production processing, such as X-ray, electromagnetic mid-sagittal articulography (EMA) \cite{wrench2000multichannel}, Magnetic Resonance Imaging (MRI) \cite{narayanan2014real} and ultrasound \cite{stone2005guide}. X-ray imaging can provide good temporal resolution, however, the subjects are exposed to radiation and it is forbidden presently. As for the EMA, the data can provide motion information by measuring the motion trajectory of the tongue. Nevertheless, EMA is invasive, which makes it difficult for natural speech production recording. MRI system can capture tongue movement with good resolution, but requires summation of repetitions to get good spatial-temporal resolution \cite{narayanan2014real}. Ultrasound is attractive, as it is non-invasive and relatively easy to conduct the experiments with good temporal and spatial resolution \cite{stone2005guide}. Moreover, ultrasound equipment is much relatively cheaper.
	
	It is well known that different imaging modality contains the information which is directly related to the speech signals. These imaging modalities are incontrovertibly statistically related each other, such as recent studies suggest that one can picture the corresponding tongue motion from their voice and vice versa \cite{porras2019dnn}. Here, we would like to ask a related question: given an observable image sequences of lips, can we predict the motion of the tongue \cite{kroos2017using}. The authors in \cite{xu2020predicting} demonstrated that deep learning model\cite{krizhevsky2012imagenet,he2016deep} can reconstruct the tongue's motion from the lip images with satisfactory performance. However, there remains considerable room for improvement for the reconstruction task.
		
	Inspired by the advancements in deep learning and cross-modal mapping\cite{feng2014cross}, we explored self-supervised learning \cite{hendrycks2019using} to build the correspondence between these two modalities. In more detail, we explore a deep neural network architecture: two-stream Convolutional long short memory network for the task, in which the deep network is trained to predict tongue motions in ultrasound image sequences based on lip image sequence. Thus, the model can leverage the temporal alignment information between two modalities. Moreover, to improve the performance further, the attention mechanism is explored. To demonstrate the effectiveness of proposed method, extensive experiments are conducted. The experimental results show that our model is able to generate images that close to the real ultrasound tongue images, and results in the matching between two imaging modalities.
	
	The remaining of this paper is organized as follows: the related work is described in the next section. Section 3 presents the methodology and experimental results are given in Section 4. Conclusions and future perspectives appear in the last section.
	
		\begin{figure*}[ht]
		\centering
		\includegraphics[width=\linewidth]{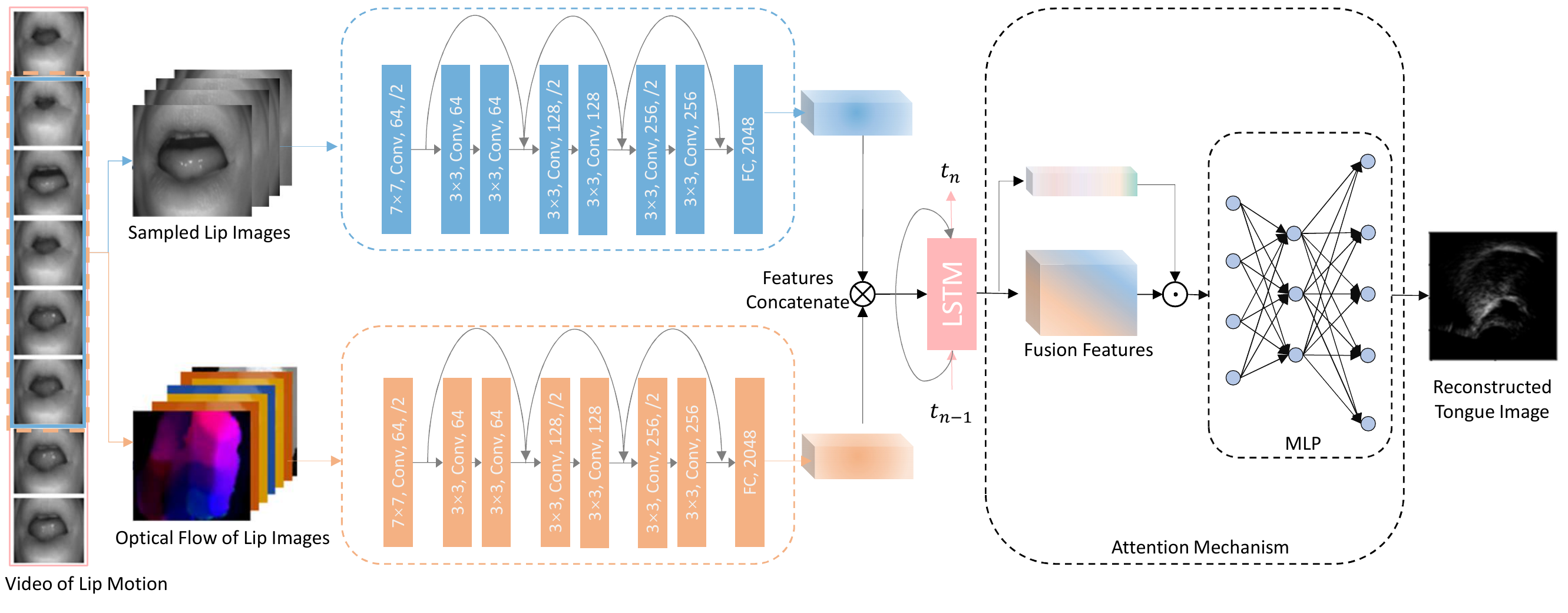}
		\caption{This figure illustrates the designed network architecture. First, the region of interest for the mouth is selected manually. One branch takes the grayscale lip images as the input, and the other inputs optical flow computed between adjacent frames. An embedding is created by concatenating the outputs of each tower. It is subsequently fed into the attention layer which consists of fully connected layers which output ultrasound tongue image.}
		\label{dif_config}
	\end{figure*}

  	 \section{Related Work}
	The task defined in this paper can be uniquely positioned in the context of two research fields: speech production study and self-supervised learning.
		
	\subsection{Speech production study}
     Measuring vocal tract’s deformation directly is difficult since the vocal tract lies within the oral cavity and is inaccessible to most instruments \cite{stone2005guide}. Indirectly imaging method has served as a valuable tool for the studies in speech production. Previous attempts focus on extract features from the images \cite{hueber2011statistical,wand2018domain,stone2020cross,gosztolya2020applying}.
    As aforementioned, these imaging modalities are incontrovertibly statistically related each other. In \cite{kroos2017using}, the authors explored the use of deep neural networks to estimate the tongue's motion from the face pictures.
    In this paper, we follow the task defined in \cite{xu2020predicting} and we aim to picture tongue's motion from the lip images, leveraging the ultrasound tongue imaging.
	
    \subsection{Self-supervised Learning}
   Self-supervised learning\cite{jing2020self,liu2019selflow} aims to build the deep model which exploits implicit labels, and the labels are free available in the data themselves. There has been an increasing interest in the self-supervised learning framework within the speech production research community \cite{eshky2019synchronising,akbari2018lip2audspec,shukla2020visually}. As we employ the temporal alignment information between the ultrasound tongue image and the lip images from stand industrial camera, our work belongs to the domain of self-supervised learning. Last two years, more attempts have been made to match the speaker identity between the face image and voice \cite{wen2019face,yoshitomi2000effect,oh2019speech2face} or audio to gesture\cite{hasegawa2018evaluation,kucherenko2019analyzing} and audio to dance\cite{zhuang2020music2dance,ferreira2021learning}. However, very few attempts have been made to build the correspondence between the ultrasound images and the lip images.
	
	\section{Methodology}
	Our goal is to generate the ultrasound tongue image sequences, which corresponds to the duration of the lip image sequences during the natural speech production process. Central to our method is the observation that the natural synchronization between the ultrasound tongue images and lip images in unlabeled video, which can serve as a form of self-supervision for learning \cite{gan2019self}. As shown in Figure 1, the architecture of our model is given, which comprised of an two-stream convolutional neural network \cite{simonyan2014two} and Long short-term memory (LSTM) network \cite{hochreiter1997long}, with attention mechanism \cite{bahdanau2015neural}. In the following of this paper, we denote our method as TS-CNN-LSTM (two-stream CNN and LSTM). The two-stream convolutional neural network takes the aforementioned lip video clip and its optical flow as inputs and compact them into a latent vector representing the visual features, which can be used to reconstruct the ultrasound tongue image using the LSTM. The details of the components will be in given subsequently.
	
	\subsection{Two stream convolutional neural network}
	As suggested in \cite{ephrat2017improved}, the instantaneous movements of the lips could be significantly disambiguated. Therefore, here we employ two streams deep model for the learning task, which including two input branches: a clip of $N$ consecutive grayscale lip image frames, and $(N-1)$ consecutive dense optical flow fields. The optical flow fields can correspond to the motion in $(u, v)$ directions for pixels of consecutive frames. Previous studies proven that optical flow can provide better performance of deep models when combined with raw input, and has even been successfully used as a stand-alone network input. Here, we crop the region of the mouth and resized the region to a fixed size of $H\times W$ (set as $96 \times 96$ in our experiments). Thus, the size of input for the first branch is $H\times W \times N$ ($N$ is the number of frames, and we found that using $N=7$ frames as input worked best). The dense optical flow is employed as the input for the second branch, which adds an additional input with the size of $H \times W \times (N-1) \times2$. Optical flow is positively influential in our experiments as well, which will be shown later.
	
	\subsection{Long short-term memory}
	To capture the temporal information from the sequence, our network architecture also contains the LSTM layer, which is one of the recurrent neural network (RNN)\cite{sak2014long} models. The core part of LSTM is the memory unit $ C_{t} $, which can be adjusted. Specifically, the input gate $ i $ encodes the information $ x_{t} $ input at the current moment and the hidden unit $ h_{t-1} $ at the previous moment, and then recognizes the information that can change the memory unit. In addition, the forget gate $ f $ can control whether to forget the information of the memory unit $ c_{t-1} $ at the previous moment. Finally, the input gate $ o $ adjusts how many proportions of memory unit information can pass and then generate hidden units $ h_{t} $. Given a sequence input $ x=\left\{x_{1},x_{2},\cdots, x_{N}  \right\} $,
	LSTM can learn the relation between the input sequences (the lip image sequences), and the updated rules are shown in the following formula:
	
	\begin{equation}
	i_{t}=\sigma(W_{xi} \times x_{t})+W_{hi}h_{t-1}+b_{i}
	\end{equation}
	
	\begin{equation}
	f_{t}=\sigma(W_{xf}\times x_{t})+W_{hf} \times h_{t-1}+b_{f}
	\end{equation}
	
	\begin{equation}
	o_{t}=\sigma(W_{xo}\times x_{t})+W_{ho} \times h_{t-1}+b_{o}
	\end{equation}
	
	\begin{equation}
	g_{t}=tanh(W_{xg} \times x_{t})+W_{hg}\times h_{t-1}+b_{g}
	\end{equation}
	
	\begin{equation}
	c_{t}=f_{t}\times c_{t-1}+i_{t}\times g_{t}
	\end{equation}
	
	\begin{equation}
	h_{t}=O_{t}\times tanh(c_{t})
	\end{equation}
	where $ x_{t} $ is a fixed-length input vector representation. $ W_{ij} $ is the weight parameter representation that links different layers. LSTM is widely-used for sequential modeling.

		
	\subsection{Attention mechanism}
	Recent studies suggested that: the performance of deep learning can be heavily decreased when the entire input features is squashed to a fixed length vector. Attention mechanism can be used to alleviate this issue, which is proposed in \cite{bahdanau2015neural}, which is a brain-like information processing mechanism based on human visual characteristics research. It can dynamically assign different weights to lip video based on the output of the TS-CNN-LSTM network at different times. We implement the attention mechanism as an additional sigmoid layer with one node output, and the propose is force the LSTM network to focus more on the frame in which the lip's motion events occur. The predicted attention factor $F_{att}(t)$ at the $t_{th}$ clips indicates the importance of each fusion feature for the final ultrasound tongue image reconstruction task: 

    \begin{equation}
    F_{att}(t)=\sigma(W_{att}*Y_{t}+b_{att}),
    \end{equation}
    here $Y_{t}$ is the output features for the LSTM netwrok, at the $t_{th}$ video clips. $\sigma$ is the sigmoid function. $W_{att}$ and $b_{att}$ denote the weights and bias of the attention mechanism. Then, the importance of each fusion feature is multiplied with the LSTM network's outputs to ensure the pipeline more deterministic:
    
    \begin{equation}
    Y_{LSTM}^{`}(t)=F_{att})(t)Y_{LSTM}(t),
    \end{equation}
    where $Y_{LSTM}(t)$ denotes the result generated from the hidden layers of the LSTM. The weighted feature for final ultrasound tongue image reconstruction is denoted by $Y_{LSTM}^{`}(t)$.
	
    During this process, the importance of some certain frames will be highlighted and the reconstruction performance will be improved.
	\section{Experiments Results}
	
	\subsection{Dataset and implementation details}
	Following \cite{xu2020predicting}, the 2010 Silent Speech Challenge (SSC) data is used for our experiments \cite{ji2018updating}. In this dataset, the ultrasound tongue images and lip videos are synchronously recorded at a speed of 60 frames per second. We randomly selected 100,000 sequence pairs from the dataset and 60\% of the data were used in training, 20\% are used for validation and the remaining 20\% were used for the evaluation.
	We implemented the network using the Keras library \footnote{https://keras.io/} built on top of TensorFlow \footnote{https://www.tensorflow.org/}. Before each activation layer, we perform the Batch Normalization process\cite{ioffe2015batch}. Dropout\cite{srivastava2014dropout} is used to prevent overfitting, and the dropout rate is set as 0.25 after convolutional layers and 0.5 after fully connected ones. The Leaky ReLU\cite{xu2015empirical} activation function is used as the non-linear activation function in all layers but the last two. We use Adam optimizer\cite{kingma2014adam} with an initial learning rate of $10*e^{-4}$. The batch size is set as 32 and stop training is conducted when the validation loss stops decreasing (around 50 epochs). We trained network using the back-propagation with mean squared error (MSE) loss. The ConvLSTM architecture proposed in \cite{xu2020predicting} is used for the quantitatively comparison. 
	
	\subsection{Performance evaluation} 
	\subsubsection{Subjective evaluation}
	Figure 2 is given to visualize the reconstructed frames, in which both the predicted ultrasound tongue image sequences and the real ones are given. As shown in the figure, our results are promising enough to indicate that reconstructing ultrasound tongue images is a feasible task, leveraging the lip image sequence as the input. The predicted contains the realistic tongue shapes. To further evaluate the reconstruction performance between different approaches,  we randomly chose 40 generated frames from the test set of the speaker. 12 subjects (4 females, 8 males; 18–21 years old; the author is not included) are asked to score the similarity between the generated ones and the original. The range of the similarity score is 1-5 (higher is better). The Table 1 shows the average similarity scores for the tested approaches. The higher score means the better performance. In general, the proposed method can provided preferred performance, compared to the ConvLSTM network.

\begin{figure}[ht]
	\centering
	\includegraphics[width=\linewidth]{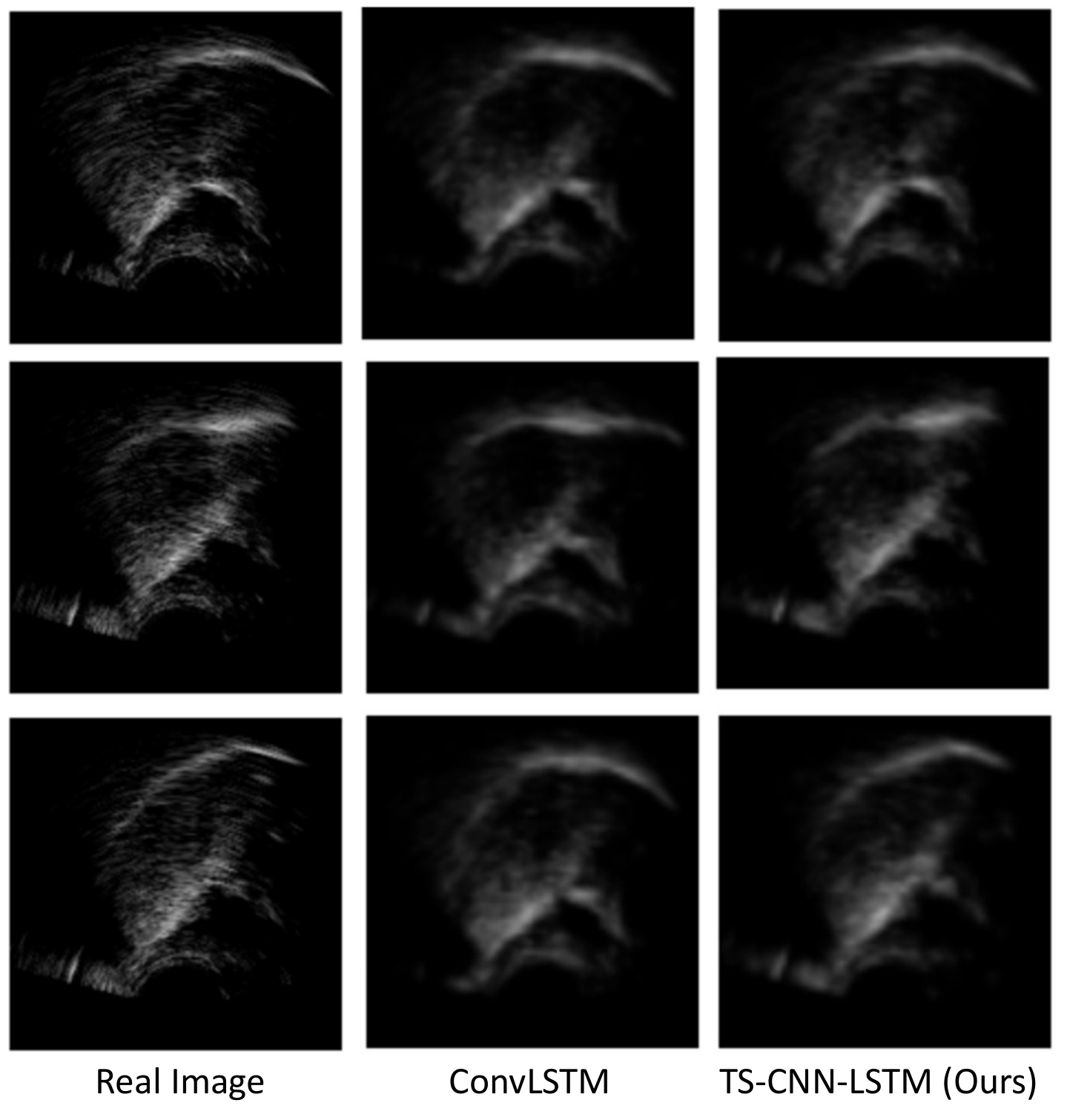}
	\caption{This figure illustrates the reconstructed ultrasound tongue images using ConvLSTM and TS-CNN-LSTM (with attention).}
	\label{rec}
\end{figure}
	
	\begin{table}[h]
		\centering 
		\caption{The similarity score for the user study from 12 subjects. Our results are given by averaging the scores for different approaches. Higher score means better performance.}
		\begin{tabular}{l|c}
			\toprule
			\textbf{Method} &  \textbf{Similarity score}  \\ 
			\midrule
			ConvLSTM\cite{xu2020predicting} & 2.67 $\pm$ 0.56 \\
			
			TS-CNN-LSTM (w/o attention) & 2.84 $\pm$ 0.49 \\
			
			TS-CNN-LSTM (Ours) & \textbf{3.01 $\pm$ 0.33}  \\
			\bottomrule
		\end{tabular}
	\end{table}

\subsubsection{Objective evaluation}
Except for the subjective evaluation, we also conduct quantitative evaluation of the learning model. To measure the naturalness of the reconstructed ultrasound images, three metrics were chosen. The first two metrics, structural similarity index (SSIM) and complex wavelet structural similarity index (CW-SSIM) \cite{sampat2009complex} are calculated over each frame from the original image sequences and the predicted ones. For the third metric, we used the mean sum of distances (MSD) to compare the contour manually extracted from the real ultrasound tongue image and the reconstructed ones. Table 2 presents the results. According to these objective experiments, all measures have shown the advantage of proposed TS-CNN-LSTM method, instead of ConvLSTM. Moreover, attention mechanism consistently improves the prediction performance.

\begin{table}[h]
	\centering 
	\caption{Quantitative comparison between different experiments settings. Higher SSIM, CW-SSIM and lower MSD means better performance. In the table, attention mechanism is denoted as "AT"}
	\begin{tabular}{@{}l|c|ccc@{}}
		\toprule
		\textbf{Method} & \textbf{AT}& \textbf{SSIM}$\uparrow$ & \textbf{CW-SSIM}$\uparrow$ & \textbf{MSD}$\downarrow$   \\ 
		\midrule
		ConvLSTM\cite{xu2020predicting} &  &0.697 & 0.719 & 5.263\\
	
		ConvLSTM\cite{xu2020predicting} & \checkmark & 0.699 & 0.728 & 5.234\\
		
		TS-CNN-LSTM & &0.711 & 0.761 & 5.056\\
	
		TS-CNN-LSTM (Ours) & \checkmark & \textbf{0.728} &  \textbf{0.765} & \textbf{4.953}\\
		\bottomrule
	\end{tabular}
\end{table}
\subsection{Ablation study}
We conducted a few ablation studies to understand the key components in our model. We compare our full model with (i) a model using only raw lip image sequences; (ii) a model with no attention mechanism; (iii) a model using raw data and optical flow as the input, but without attention mechanism. Table 3 shows the results of this analysis. Our analysis shows that adding optical flow provide most of the information needed for reconstructing speech, while the attention mechanism give slightly better results.

\begin{table}[h]
	\centering 
	\caption{Results of ablation analysis. Adding optical flow can provide most of the information needed, and attention mechanism gives slightly better results. In the table, optical flow and attention mechanism are denoted as ``OF'' and "AT", respectively.}
	\begin{tabular}{@{}l|c|c@{}}
		\toprule
		\textbf{Input} & \textbf{SSIM} & \textbf{CW-SSIM} \\ 
		\midrule
		Raw Images (with TS-CNN-LSTM) & 0.683 & 0.706\\
		
		Ours (w/o OF) &0.691 & 0.712 \\
		
		Ours (w/o AT) & 0.711 & 0.761 \\
		
		Ours & \textbf{0.728} &  \textbf{0.765} \\
		\bottomrule
	\end{tabular}
\end{table}

	\section{Conclusion}
	We aim to reconstruct the tongue image from the simultaneously recorded lip motion using the TS-CNN-LSTM network, and the proposed method is seen to achieve better reconstruction results through subjective and objective evaluation compared to ConvLSTM architecture. As aforementioned, our method might be useful for speech production studies, articulatory motion synthesis and for pronunciation training. It is worthwhile to notice that our method is speaker-dependent, and a new model needs to be fine-tuned for each new speaker. Achieving speaker-independent ultrasound tongue image reconstruction is a non-trivial task, and will be explored in our future work.

	\bibliographystyle{ACM-Reference-Format}
	\bibliography{bibliography}


\begin{thebibliography}{41}


\ifx \showCODEN    \undefined \def \showCODEN     #1{\unskip}     \fi
\ifx \showDOI      \undefined \def \showDOI       #1{#1}\fi
\ifx \showISBNx    \undefined \def \showISBNx     #1{\unskip}     \fi
\ifx \showISBNxiii \undefined \def \showISBNxiii  #1{\unskip}     \fi
\ifx \showISSN     \undefined \def \showISSN      #1{\unskip}     \fi
\ifx \showLCCN     \undefined \def \showLCCN      #1{\unskip}     \fi
\ifx \shownote     \undefined \def \shownote      #1{#1}          \fi
\ifx \showarticletitle \undefined \def \showarticletitle #1{#1}   \fi
\ifx \showURL      \undefined \def \showURL       {\relax}        \fi
\providecommand\bibfield[2]{#2}
\providecommand\bibinfo[2]{#2}
\providecommand\natexlab[1]{#1}
\providecommand\showeprint[2][]{arXiv:#2}

\bibitem[\protect\citeauthoryear{Akbari, Arora, Cao, and Mesgarani}{Akbari
  et~al\mbox{.}}{2018}]%
        {akbari2018lip2audspec}
\bibfield{author}{\bibinfo{person}{Hassan Akbari}, \bibinfo{person}{Himani
  Arora}, \bibinfo{person}{Liangliang Cao}, {and} \bibinfo{person}{Nima
  Mesgarani}.} \bibinfo{year}{2018}\natexlab{}.
\newblock \showarticletitle{Lip2audspec: Speech reconstruction from silent lip
  movements video}. In \bibinfo{booktitle}{\emph{2018 IEEE International
  Conference on Acoustics, Speech and Signal Processing (ICASSP)}}. IEEE,
  \bibinfo{pages}{2516--2520}.
\newblock


\bibitem[\protect\citeauthoryear{Bahdanau, Cho, and Bengio}{Bahdanau
  et~al\mbox{.}}{2015}]%
        {bahdanau2015neural}
\bibfield{author}{\bibinfo{person}{Dzmitry Bahdanau},
  \bibinfo{person}{Kyung~Hyun Cho}, {and} \bibinfo{person}{Yoshua Bengio}.}
  \bibinfo{year}{2015}\natexlab{}.
\newblock \showarticletitle{Neural machine translation by jointly learning to
  align and translate}. In \bibinfo{booktitle}{\emph{3rd International
  Conference on Learning Representations, ICLR 2015}}.
\newblock


\bibitem[\protect\citeauthoryear{Denby, Schultz, Honda, Hueber, Gilbert, and
  Brumberg}{Denby et~al\mbox{.}}{2010}]%
        {denby2010silent}
\bibfield{author}{\bibinfo{person}{Bruce Denby}, \bibinfo{person}{Tanja
  Schultz}, \bibinfo{person}{Kiyoshi Honda}, \bibinfo{person}{Thomas Hueber},
  \bibinfo{person}{Jim~M Gilbert}, {and} \bibinfo{person}{Jonathan~S
  Brumberg}.} \bibinfo{year}{2010}\natexlab{}.
\newblock \showarticletitle{Silent speech interfaces}.
\newblock \bibinfo{journal}{\emph{Speech Communication}} \bibinfo{volume}{52},
  \bibinfo{number}{4} (\bibinfo{year}{2010}), \bibinfo{pages}{270--287}.
\newblock


\bibitem[\protect\citeauthoryear{Ding, Liberatore, Sonsaat, Lu{\v{c}}i{\'c},
  Silpachai, Zhao, Chukharev-Hudilainen, Levis, and Gutierrez-Osuna}{Ding
  et~al\mbox{.}}{2019}]%
        {ding2019golden}
\bibfield{author}{\bibinfo{person}{Shaojin Ding}, \bibinfo{person}{Christopher
  Liberatore}, \bibinfo{person}{Sinem Sonsaat}, \bibinfo{person}{Ivana
  Lu{\v{c}}i{\'c}}, \bibinfo{person}{Alif Silpachai}, \bibinfo{person}{Guanlong
  Zhao}, \bibinfo{person}{Evgeny Chukharev-Hudilainen}, \bibinfo{person}{John
  Levis}, {and} \bibinfo{person}{Ricardo Gutierrez-Osuna}.}
  \bibinfo{year}{2019}\natexlab{}.
\newblock \showarticletitle{Golden speaker builder--An interactive tool for
  pronunciation training}.
\newblock \bibinfo{journal}{\emph{Speech Communication}}  \bibinfo{volume}{115}
  (\bibinfo{year}{2019}), \bibinfo{pages}{51--66}.
\newblock


\bibitem[\protect\citeauthoryear{Ephrat, Halperin, and Peleg}{Ephrat
  et~al\mbox{.}}{2017}]%
        {ephrat2017improved}
\bibfield{author}{\bibinfo{person}{Ariel Ephrat}, \bibinfo{person}{Tavi
  Halperin}, {and} \bibinfo{person}{Shmuel Peleg}.}
  \bibinfo{year}{2017}\natexlab{}.
\newblock \showarticletitle{Improved speech reconstruction from silent video}.
  In \bibinfo{booktitle}{\emph{Proceedings of the IEEE International Conference
  on Computer Vision Workshops}}. \bibinfo{pages}{455--462}.
\newblock


\bibitem[\protect\citeauthoryear{Eshky, Ribeiro, Richmond, and Renals}{Eshky
  et~al\mbox{.}}{2019}]%
        {eshky2019synchronising}
\bibfield{author}{\bibinfo{person}{Aciel Eshky}, \bibinfo{person}{Manuel~Sam
  Ribeiro}, \bibinfo{person}{Korin Richmond}, {and} \bibinfo{person}{Steve
  Renals}.} \bibinfo{year}{2019}\natexlab{}.
\newblock \showarticletitle{Synchronising audio and ultrasound by learning
  cross-modal embeddings}.
\newblock  (\bibinfo{year}{2019}).
\newblock


\bibitem[\protect\citeauthoryear{Feng, Wang, and Li}{Feng
  et~al\mbox{.}}{2014}]%
        {feng2014cross}
\bibfield{author}{\bibinfo{person}{Fangxiang Feng}, \bibinfo{person}{Xiaojie
  Wang}, {and} \bibinfo{person}{Ruifan Li}.} \bibinfo{year}{2014}\natexlab{}.
\newblock \showarticletitle{Cross-modal retrieval with correspondence
  autoencoder}. In \bibinfo{booktitle}{\emph{Proceedings of the 22nd ACM
  international conference on Multimedia}}. \bibinfo{pages}{7--16}.
\newblock


\bibitem[\protect\citeauthoryear{Ferreira, Coutinho, Gomes, Neto, Azevedo,
  Martins, and Nascimento}{Ferreira et~al\mbox{.}}{2021}]%
        {ferreira2021learning}
\bibfield{author}{\bibinfo{person}{Joao~P Ferreira}, \bibinfo{person}{Thiago~M
  Coutinho}, \bibinfo{person}{Thiago~L Gomes}, \bibinfo{person}{Jos{\'e}~F
  Neto}, \bibinfo{person}{Rafael Azevedo}, \bibinfo{person}{Renato Martins},
  {and} \bibinfo{person}{Erickson~R Nascimento}.}
  \bibinfo{year}{2021}\natexlab{}.
\newblock \showarticletitle{Learning to dance: A graph convolutional
  adversarial network to generate realistic dance motions from audio}.
\newblock \bibinfo{journal}{\emph{Computers \& Graphics}}  \bibinfo{volume}{94}
  (\bibinfo{year}{2021}), \bibinfo{pages}{11--21}.
\newblock


\bibitem[\protect\citeauthoryear{Gan, Zhao, Chen, Cox, and Torralba}{Gan
  et~al\mbox{.}}{2019}]%
        {gan2019self}
\bibfield{author}{\bibinfo{person}{Chuang Gan}, \bibinfo{person}{Hang Zhao},
  \bibinfo{person}{Peihao Chen}, \bibinfo{person}{David Cox}, {and}
  \bibinfo{person}{Antonio Torralba}.} \bibinfo{year}{2019}\natexlab{}.
\newblock \showarticletitle{Self-supervised moving vehicle tracking with stereo
  sound}. In \bibinfo{booktitle}{\emph{Proceedings of the IEEE International
  Conference on Computer Vision}}. \bibinfo{pages}{7053--7062}.
\newblock


\bibitem[\protect\citeauthoryear{Gosztolya, Gr{\'o}sz, T{\'o}th, Mark{\'o}, and
  Csap{\'o}}{Gosztolya et~al\mbox{.}}{2020}]%
        {gosztolya2020applying}
\bibfield{author}{\bibinfo{person}{G{\'a}bor Gosztolya},
  \bibinfo{person}{Tam{\'a}s Gr{\'o}sz}, \bibinfo{person}{L{\'a}szl{\'o}
  T{\'o}th}, \bibinfo{person}{Alexandra Mark{\'o}}, {and}
  \bibinfo{person}{Tam{\'a}s~G{\'a}bor Csap{\'o}}.}
  \bibinfo{year}{2020}\natexlab{}.
\newblock \showarticletitle{Applying dnn adaptation to reduce the session
  dependency of ultrasound tongue imaging-based silent speech interfaces}.
\newblock \bibinfo{journal}{\emph{Acta Polytechnica Hungarica}}
  \bibinfo{volume}{17}, \bibinfo{number}{7} (\bibinfo{year}{2020}),
  \bibinfo{pages}{109--128}.
\newblock


\bibitem[\protect\citeauthoryear{Hasegawa, Kaneko, Shirakawa, Sakuta, and
  Sumi}{Hasegawa et~al\mbox{.}}{2018}]%
        {hasegawa2018evaluation}
\bibfield{author}{\bibinfo{person}{Dai Hasegawa}, \bibinfo{person}{Naoshi
  Kaneko}, \bibinfo{person}{Shinichi Shirakawa}, \bibinfo{person}{Hiroshi
  Sakuta}, {and} \bibinfo{person}{Kazuhiko Sumi}.}
  \bibinfo{year}{2018}\natexlab{}.
\newblock \showarticletitle{Evaluation of speech-to-gesture generation using
  bi-directional LSTM network}. In \bibinfo{booktitle}{\emph{Proceedings of the
  18th International Conference on Intelligent Virtual Agents}}.
  \bibinfo{pages}{79--86}.
\newblock


\bibitem[\protect\citeauthoryear{He, Zhang, Ren, and Sun}{He
  et~al\mbox{.}}{2016}]%
        {he2016deep}
\bibfield{author}{\bibinfo{person}{Kaiming He}, \bibinfo{person}{Xiangyu
  Zhang}, \bibinfo{person}{Shaoqing Ren}, {and} \bibinfo{person}{Jian Sun}.}
  \bibinfo{year}{2016}\natexlab{}.
\newblock \showarticletitle{Deep residual learning for image recognition}. In
  \bibinfo{booktitle}{\emph{Proceedings of the IEEE conference on computer
  vision and pattern recognition}}. \bibinfo{pages}{770--778}.
\newblock


\bibitem[\protect\citeauthoryear{Hendrycks, Mazeika, Kadavath, and
  Song}{Hendrycks et~al\mbox{.}}{2019}]%
        {hendrycks2019using}
\bibfield{author}{\bibinfo{person}{Dan Hendrycks}, \bibinfo{person}{Mantas
  Mazeika}, \bibinfo{person}{Saurav Kadavath}, {and} \bibinfo{person}{Dawn
  Song}.} \bibinfo{year}{2019}\natexlab{}.
\newblock \showarticletitle{Using self-supervised learning can improve model
  robustness and uncertainty}. In \bibinfo{booktitle}{\emph{Advances in Neural
  Information Processing Systems}}. \bibinfo{pages}{15663--15674}.
\newblock


\bibitem[\protect\citeauthoryear{Hochreiter and Schmidhuber}{Hochreiter and
  Schmidhuber}{1997}]%
        {hochreiter1997long}
\bibfield{author}{\bibinfo{person}{Sepp Hochreiter} {and}
  \bibinfo{person}{J{\"u}rgen Schmidhuber}.} \bibinfo{year}{1997}\natexlab{}.
\newblock \showarticletitle{Long short-term memory}.
\newblock \bibinfo{journal}{\emph{Neural computation}} \bibinfo{volume}{9},
  \bibinfo{number}{8} (\bibinfo{year}{1997}), \bibinfo{pages}{1735--1780}.
\newblock


\bibitem[\protect\citeauthoryear{Hueber, Benaroya, Denby, and Chollet}{Hueber
  et~al\mbox{.}}{2011}]%
        {hueber2011statistical}
\bibfield{author}{\bibinfo{person}{Thomas Hueber},
  \bibinfo{person}{Elie-Laurent Benaroya}, \bibinfo{person}{Bruce Denby}, {and}
  \bibinfo{person}{G{\'e}rard Chollet}.} \bibinfo{year}{2011}\natexlab{}.
\newblock \showarticletitle{Statistical mapping between articulatory and
  acoustic data for an ultrasound-based silent speech interface}. In
  \bibinfo{booktitle}{\emph{Twelfth Annual Conference of the International
  Speech Communication Association}}.
\newblock


\bibitem[\protect\citeauthoryear{Ioffe and Szegedy}{Ioffe and Szegedy}{2015}]%
        {ioffe2015batch}
\bibfield{author}{\bibinfo{person}{Sergey Ioffe} {and}
  \bibinfo{person}{Christian Szegedy}.} \bibinfo{year}{2015}\natexlab{}.
\newblock \showarticletitle{Batch normalization: Accelerating deep network
  training by reducing internal covariate shift}. In
  \bibinfo{booktitle}{\emph{International conference on machine learning}}.
  PMLR, \bibinfo{pages}{448--456}.
\newblock


\bibitem[\protect\citeauthoryear{Ji, Liu, Wang, Liu, Niu, and Denby}{Ji
  et~al\mbox{.}}{2018}]%
        {ji2018updating}
\bibfield{author}{\bibinfo{person}{Yan Ji}, \bibinfo{person}{Licheng Liu},
  \bibinfo{person}{Hongcui Wang}, \bibinfo{person}{Zhilei Liu},
  \bibinfo{person}{Zhibin Niu}, {and} \bibinfo{person}{Bruce Denby}.}
  \bibinfo{year}{2018}\natexlab{}.
\newblock \showarticletitle{Updating the silent speech challenge benchmark with
  deep learning}.
\newblock \bibinfo{journal}{\emph{Speech Communication}}  \bibinfo{volume}{98}
  (\bibinfo{year}{2018}), \bibinfo{pages}{42--50}.
\newblock


\bibitem[\protect\citeauthoryear{Jing and Tian}{Jing and Tian}{2020}]%
        {jing2020self}
\bibfield{author}{\bibinfo{person}{Longlong Jing} {and} \bibinfo{person}{Yingli
  Tian}.} \bibinfo{year}{2020}\natexlab{}.
\newblock \showarticletitle{Self-supervised visual feature learning with deep
  neural networks: A survey}.
\newblock \bibinfo{journal}{\emph{IEEE Transactions on Pattern Analysis and
  Machine Intelligence}} (\bibinfo{year}{2020}).
\newblock


\bibitem[\protect\citeauthoryear{Kingma and Ba}{Kingma and Ba}{2014}]%
        {kingma2014adam}
\bibfield{author}{\bibinfo{person}{Diederik~P Kingma} {and}
  \bibinfo{person}{Jimmy Ba}.} \bibinfo{year}{2014}\natexlab{}.
\newblock \showarticletitle{Adam: A method for stochastic optimization}.
\newblock \bibinfo{journal}{\emph{arXiv preprint arXiv:1412.6980}}
  (\bibinfo{year}{2014}).
\newblock


\bibitem[\protect\citeauthoryear{Krizhevsky, Sutskever, and Hinton}{Krizhevsky
  et~al\mbox{.}}{2012}]%
        {krizhevsky2012imagenet}
\bibfield{author}{\bibinfo{person}{Alex Krizhevsky}, \bibinfo{person}{Ilya
  Sutskever}, {and} \bibinfo{person}{Geoffrey~E Hinton}.}
  \bibinfo{year}{2012}\natexlab{}.
\newblock \showarticletitle{Imagenet classification with deep convolutional
  neural networks}.
\newblock \bibinfo{journal}{\emph{Advances in neural information processing
  systems}}  \bibinfo{volume}{25} (\bibinfo{year}{2012}),
  \bibinfo{pages}{1097--1105}.
\newblock


\bibitem[\protect\citeauthoryear{Kroos, Bundgaard-Nielsen, Best, and
  Plumbley}{Kroos et~al\mbox{.}}{2017}]%
        {kroos2017using}
\bibfield{author}{\bibinfo{person}{Christian Kroos}, \bibinfo{person}{Rikke~L
  Bundgaard-Nielsen}, \bibinfo{person}{Catherine~T Best}, {and}
  \bibinfo{person}{Mark Plumbley}.} \bibinfo{year}{2017}\natexlab{}.
\newblock \showarticletitle{Using deep neural networks to estimate tongue
  movements from speech face motion}.
\newblock \bibinfo{journal}{\emph{Proceedings of AVSP 2017}}
  (\bibinfo{year}{2017}).
\newblock


\bibitem[\protect\citeauthoryear{Kucherenko, Hasegawa, Henter, Kaneko, and
  Kjellstr{\"o}m}{Kucherenko et~al\mbox{.}}{2019}]%
        {kucherenko2019analyzing}
\bibfield{author}{\bibinfo{person}{Taras Kucherenko}, \bibinfo{person}{Dai
  Hasegawa}, \bibinfo{person}{Gustav~Eje Henter}, \bibinfo{person}{Naoshi
  Kaneko}, {and} \bibinfo{person}{Hedvig Kjellstr{\"o}m}.}
  \bibinfo{year}{2019}\natexlab{}.
\newblock \showarticletitle{Analyzing input and output representations for
  speech-driven gesture generation}. In \bibinfo{booktitle}{\emph{Proceedings
  of the 19th ACM International Conference on Intelligent Virtual Agents}}.
  \bibinfo{pages}{97--104}.
\newblock


\bibitem[\protect\citeauthoryear{Liu, Lyu, King, and Xu}{Liu
  et~al\mbox{.}}{2019}]%
        {liu2019selflow}
\bibfield{author}{\bibinfo{person}{Pengpeng Liu}, \bibinfo{person}{Michael
  Lyu}, \bibinfo{person}{Irwin King}, {and} \bibinfo{person}{Jia Xu}.}
  \bibinfo{year}{2019}\natexlab{}.
\newblock \showarticletitle{Selflow: Self-supervised learning of optical flow}.
  In \bibinfo{booktitle}{\emph{Proceedings of the IEEE/CVF Conference on
  Computer Vision and Pattern Recognition}}. \bibinfo{pages}{4571--4580}.
\newblock


\bibitem[\protect\citeauthoryear{Narayanan, Toutios, Ramanarayanan, Lammert,
  Kim, Lee, Nayak, Kim, Zhu, Goldstein, et~al\mbox{.}}{Narayanan
  et~al\mbox{.}}{2014}]%
        {narayanan2014real}
\bibfield{author}{\bibinfo{person}{Shrikanth Narayanan},
  \bibinfo{person}{Asterios Toutios}, \bibinfo{person}{Vikram Ramanarayanan},
  \bibinfo{person}{Adam Lammert}, \bibinfo{person}{Jangwon Kim},
  \bibinfo{person}{Sungbok Lee}, \bibinfo{person}{Krishna Nayak},
  \bibinfo{person}{Yoon-Chul Kim}, \bibinfo{person}{Yinghua Zhu},
  \bibinfo{person}{Louis Goldstein}, {et~al\mbox{.}}}
  \bibinfo{year}{2014}\natexlab{}.
\newblock \showarticletitle{Real-time magnetic resonance imaging and
  electromagnetic articulography database for speech production research (TC)}.
\newblock \bibinfo{journal}{\emph{The Journal of the Acoustical Society of
  America}} \bibinfo{volume}{136}, \bibinfo{number}{3} (\bibinfo{year}{2014}),
  \bibinfo{pages}{1307--1311}.
\newblock


\bibitem[\protect\citeauthoryear{Oh, Dekel, Kim, Mosseri, Freeman, Rubinstein,
  and Matusik}{Oh et~al\mbox{.}}{2019}]%
        {oh2019speech2face}
\bibfield{author}{\bibinfo{person}{Tae-Hyun Oh}, \bibinfo{person}{Tali Dekel},
  \bibinfo{person}{Changil Kim}, \bibinfo{person}{Inbar Mosseri},
  \bibinfo{person}{William~T Freeman}, \bibinfo{person}{Michael Rubinstein},
  {and} \bibinfo{person}{Wojciech Matusik}.} \bibinfo{year}{2019}\natexlab{}.
\newblock \showarticletitle{Speech2face: Learning the face behind a voice}. In
  \bibinfo{booktitle}{\emph{Proceedings of the IEEE/CVF Conference on Computer
  Vision and Pattern Recognition}}. \bibinfo{pages}{7539--7548}.
\newblock


\bibitem[\protect\citeauthoryear{Porras, Sep{\'u}lveda-Sep{\'u}lveda, and
  Csap{\'o}}{Porras et~al\mbox{.}}{2019}]%
        {porras2019dnn}
\bibfield{author}{\bibinfo{person}{Dagoberto Porras},
  \bibinfo{person}{Alexander Sep{\'u}lveda-Sep{\'u}lveda}, {and}
  \bibinfo{person}{Tam{\'a}s~G{\'a}bor Csap{\'o}}.}
  \bibinfo{year}{2019}\natexlab{}.
\newblock \showarticletitle{DNN-based Acoustic-to-Articulatory Inversion using
  Ultrasound Tongue Imaging}. In \bibinfo{booktitle}{\emph{2019 International
  Joint Conference on Neural Networks (IJCNN)}}. IEEE, \bibinfo{pages}{1--8}.
\newblock


\bibitem[\protect\citeauthoryear{Sak, Senior, and Beaufays}{Sak
  et~al\mbox{.}}{2014}]%
        {sak2014long}
\bibfield{author}{\bibinfo{person}{Hasim Sak}, \bibinfo{person}{Andrew Senior},
  {and} \bibinfo{person}{Fran{\c{c}}oise Beaufays}.}
  \bibinfo{year}{2014}\natexlab{}.
\newblock \showarticletitle{Long Short-Term Memory Recurrent Neural Network
  Architectures for Large Scale Acoustic Modeling}.
\newblock  (\bibinfo{year}{2014}).
\newblock


\bibitem[\protect\citeauthoryear{Sampat, Wang, Gupta, Bovik, and Markey}{Sampat
  et~al\mbox{.}}{2009}]%
        {sampat2009complex}
\bibfield{author}{\bibinfo{person}{Mehul~P Sampat}, \bibinfo{person}{Zhou
  Wang}, \bibinfo{person}{Shalini Gupta}, \bibinfo{person}{Alan~Conrad Bovik},
  {and} \bibinfo{person}{Mia~K Markey}.} \bibinfo{year}{2009}\natexlab{}.
\newblock \showarticletitle{Complex wavelet structural similarity: A new image
  similarity index}.
\newblock \bibinfo{journal}{\emph{IEEE transactions on image processing}}
  \bibinfo{volume}{18}, \bibinfo{number}{11} (\bibinfo{year}{2009}),
  \bibinfo{pages}{2385--2401}.
\newblock


\bibitem[\protect\citeauthoryear{Schultz, Wand, Hueber, Krusienski, Herff, and
  Brumberg}{Schultz et~al\mbox{.}}{2017}]%
        {schultz2017biosignal}
\bibfield{author}{\bibinfo{person}{Tanja Schultz}, \bibinfo{person}{Michael
  Wand}, \bibinfo{person}{Thomas Hueber}, \bibinfo{person}{Dean~J Krusienski},
  \bibinfo{person}{Christian Herff}, {and} \bibinfo{person}{Jonathan~S
  Brumberg}.} \bibinfo{year}{2017}\natexlab{}.
\newblock \showarticletitle{Biosignal-based spoken communication: A survey}.
\newblock \bibinfo{journal}{\emph{IEEE/ACM Transactions on Audio, Speech, and
  Language Processing}} \bibinfo{volume}{25}, \bibinfo{number}{12}
  (\bibinfo{year}{2017}), \bibinfo{pages}{2257--2271}.
\newblock


\bibitem[\protect\citeauthoryear{Shukla, Vougioukas, Ma, Petridis, and
  Pantic}{Shukla et~al\mbox{.}}{2020}]%
        {shukla2020visually}
\bibfield{author}{\bibinfo{person}{Abhinav Shukla},
  \bibinfo{person}{Konstantinos Vougioukas}, \bibinfo{person}{Pingchuan Ma},
  \bibinfo{person}{Stavros Petridis}, {and} \bibinfo{person}{Maja Pantic}.}
  \bibinfo{year}{2020}\natexlab{}.
\newblock \showarticletitle{Visually guided self supervised learning of speech
  representations}. In \bibinfo{booktitle}{\emph{ICASSP 2020-2020 IEEE
  International Conference on Acoustics, Speech and Signal Processing
  (ICASSP)}}. IEEE, \bibinfo{pages}{6299--6303}.
\newblock


\bibitem[\protect\citeauthoryear{Simonyan and Zisserman}{Simonyan and
  Zisserman}{2014}]%
        {simonyan2014two}
\bibfield{author}{\bibinfo{person}{Karen Simonyan} {and}
  \bibinfo{person}{Andrew Zisserman}.} \bibinfo{year}{2014}\natexlab{}.
\newblock \showarticletitle{Two-stream convolutional networks for action
  recognition in videos}. In \bibinfo{booktitle}{\emph{Proceedings of the 27th
  International Conference on Neural Information Processing Systems-Volume 1}}.
  \bibinfo{pages}{568--576}.
\newblock


\bibitem[\protect\citeauthoryear{Srivastava, Hinton, Krizhevsky, Sutskever, and
  Salakhutdinov}{Srivastava et~al\mbox{.}}{2014}]%
        {srivastava2014dropout}
\bibfield{author}{\bibinfo{person}{Nitish Srivastava},
  \bibinfo{person}{Geoffrey Hinton}, \bibinfo{person}{Alex Krizhevsky},
  \bibinfo{person}{Ilya Sutskever}, {and} \bibinfo{person}{Ruslan
  Salakhutdinov}.} \bibinfo{year}{2014}\natexlab{}.
\newblock \showarticletitle{Dropout: a simple way to prevent neural networks
  from overfitting}.
\newblock \bibinfo{journal}{\emph{The journal of machine learning research}}
  \bibinfo{volume}{15}, \bibinfo{number}{1} (\bibinfo{year}{2014}),
  \bibinfo{pages}{1929--1958}.
\newblock


\bibitem[\protect\citeauthoryear{Stone}{Stone}{2005}]%
        {stone2005guide}
\bibfield{author}{\bibinfo{person}{Maureen Stone}.}
  \bibinfo{year}{2005}\natexlab{}.
\newblock \showarticletitle{A guide to analysing tongue motion from ultrasound
  images}.
\newblock \bibinfo{journal}{\emph{Clinical linguistics \& phonetics}}
  \bibinfo{volume}{19}, \bibinfo{number}{6-7} (\bibinfo{year}{2005}),
  \bibinfo{pages}{455--501}.
\newblock


\bibitem[\protect\citeauthoryear{Stone and Birkholz}{Stone and
  Birkholz}{2020}]%
        {stone2020cross}
\bibfield{author}{\bibinfo{person}{Simon Stone} {and} \bibinfo{person}{Peter
  Birkholz}.} \bibinfo{year}{2020}\natexlab{}.
\newblock \showarticletitle{Cross-speaker silent-speech command word
  recognition using electro-optical stomatography}. In
  \bibinfo{booktitle}{\emph{ICASSP 2020-2020 IEEE International Conference on
  Acoustics, Speech and Signal Processing (ICASSP)}}. IEEE,
  \bibinfo{pages}{7849--7853}.
\newblock


\bibitem[\protect\citeauthoryear{Wand, Schultz, and Schmidhuber}{Wand
  et~al\mbox{.}}{2018}]%
        {wand2018domain}
\bibfield{author}{\bibinfo{person}{Michael Wand}, \bibinfo{person}{Tanja
  Schultz}, {and} \bibinfo{person}{J{\"u}rgen Schmidhuber}.}
  \bibinfo{year}{2018}\natexlab{}.
\newblock \showarticletitle{Domain-Adversarial Training for Session Independent
  EMG-based Speech Recognition.}. In \bibinfo{booktitle}{\emph{INTERSPEECH}}.
  \bibinfo{pages}{3167--3171}.
\newblock


\bibitem[\protect\citeauthoryear{Wen, Raj, and Singh}{Wen
  et~al\mbox{.}}{2019}]%
        {wen2019face}
\bibfield{author}{\bibinfo{person}{Yandong Wen}, \bibinfo{person}{Bhiksha Raj},
  {and} \bibinfo{person}{Rita Singh}.} \bibinfo{year}{2019}\natexlab{}.
\newblock \showarticletitle{Face reconstruction from voice using generative
  adversarial networks}. In \bibinfo{booktitle}{\emph{Advances in Neural
  Information Processing Systems}}. \bibinfo{pages}{5265--5274}.
\newblock


\bibitem[\protect\citeauthoryear{Wrench}{Wrench}{2000}]%
        {wrench2000multichannel}
\bibfield{author}{\bibinfo{person}{Alan~A Wrench}.}
  \bibinfo{year}{2000}\natexlab{}.
\newblock \showarticletitle{A multichannel articulatory database and its
  application for automatic speech recognition}. In
  \bibinfo{booktitle}{\emph{In Proceedings 5 th Seminar of Speech Production}}.
  Citeseer.
\newblock


\bibitem[\protect\citeauthoryear{Xu, Wang, Chen, and Li}{Xu
  et~al\mbox{.}}{2015}]%
        {xu2015empirical}
\bibfield{author}{\bibinfo{person}{Bing Xu}, \bibinfo{person}{Naiyan Wang},
  \bibinfo{person}{Tianqi Chen}, {and} \bibinfo{person}{Mu Li}.}
  \bibinfo{year}{2015}\natexlab{}.
\newblock \showarticletitle{Empirical evaluation of rectified activations in
  convolutional network}.
\newblock \bibinfo{journal}{\emph{arXiv preprint arXiv:1505.00853}}
  (\bibinfo{year}{2015}).
\newblock


\bibitem[\protect\citeauthoryear{Xu, Zhao, Zhu, and Zhao}{Xu
  et~al\mbox{.}}{2020}]%
        {xu2020predicting}
\bibfield{author}{\bibinfo{person}{Kele Xu}, \bibinfo{person}{Jianqiao Zhao},
  \bibinfo{person}{Boqing Zhu}, {and} \bibinfo{person}{Chaojie Zhao}.}
  \bibinfo{year}{2020}\natexlab{}.
\newblock \showarticletitle{Predicting ultrasound tongue image from lip images
  using sequence to sequence learning}.
\newblock \bibinfo{journal}{\emph{The Journal of the Acoustical Society of
  America}} \bibinfo{volume}{147}, \bibinfo{number}{6} (\bibinfo{year}{2020}),
  \bibinfo{pages}{EL441--EL446}.
\newblock


\bibitem[\protect\citeauthoryear{Yoshitomi, Kim, Kawano, and Kilazoe}{Yoshitomi
  et~al\mbox{.}}{2000}]%
        {yoshitomi2000effect}
\bibfield{author}{\bibinfo{person}{Yasunari Yoshitomi},
  \bibinfo{person}{Sung-Ill Kim}, \bibinfo{person}{Takako Kawano}, {and}
  \bibinfo{person}{Tetsuro Kilazoe}.} \bibinfo{year}{2000}\natexlab{}.
\newblock \showarticletitle{Effect of sensor fusion for recognition of
  emotional states using voice, face image and thermal image of face}. In
  \bibinfo{booktitle}{\emph{Proceedings 9th IEEE International Workshop on
  Robot and Human Interactive Communication. IEEE RO-MAN 2000 (Cat. No.
  00TH8499)}}. IEEE, \bibinfo{pages}{178--183}.
\newblock


\bibitem[\protect\citeauthoryear{Zhuang, Wang, Xia, Chai, and Wang}{Zhuang
  et~al\mbox{.}}{2020}]%
        {zhuang2020music2dance}
\bibfield{author}{\bibinfo{person}{Wenlin Zhuang}, \bibinfo{person}{Congyi
  Wang}, \bibinfo{person}{Siyu Xia}, \bibinfo{person}{Jinxiang Chai}, {and}
  \bibinfo{person}{Yangang Wang}.} \bibinfo{year}{2020}\natexlab{}.
\newblock \showarticletitle{Music2dance: Music-driven dance generation using
  wavenet}.
\newblock \bibinfo{journal}{\emph{arXiv preprint arXiv:2002.03761}}
  \bibinfo{volume}{1}, \bibinfo{number}{2} (\bibinfo{year}{2020}),
  \bibinfo{pages}{7}.
\newblock


\end{thebibliography}
	
\end{document}